\documentclass[11pt,tightenlines,eqsecnum,floats,aps,amsmath,amssymb,nofootinbib,prd,shownopacs,floatfix]{revtex4}

\usepackage{graphicx}
\usepackage{epstopdf}
\usepackage{latexsym}
\usepackage{amssymb}
\usepackage{amsmath}
\usepackage{color}
\usepackage{mathrsfs}
\usepackage{xparse}
\usepackage{float}
\usepackage{mathtools}
\usepackage{multirow}
\usepackage[center]{subfigure}

\usepackage{xcolor}
\allowdisplaybreaks

\begin{document}
\renewcommand\arraystretch{2}
 \newcommand{\bq}{\begin{equation}}
 \newcommand{\eq}{\end{equation}}
 \newcommand{\bqn}{\begin{eqnarray}}
 \newcommand{\eqn}{\end{eqnarray}}
 \newcommand{\nb}{\nonumber}
 \newcommand{\cb}{\color{black}}
    \newcommand{\cc}{\color{cyan}}
     \newcommand{\lb}{\label}
        \newcommand{\cm}{\color{magenta}}
\newcommand{\rc}{\rho^{\scriptscriptstyle{\mathrm{I}}}_c}
\newcommand{\rd}{\rho^{\scriptscriptstyle{\mathrm{II}}}_c}
\NewDocumentCommand{\evalat}{sO{\big}mm}{%
  \IfBooleanTF{#1}
   {\mleft. #3 \mright|_{#4}}
   {#3#2|_{#4}}%
}
\newcommand{\PRL}{Phys. Rev. Lett.}
\newcommand{\PL}{Phys. Lett.}
\newcommand{\PR}{Phys. Rev.}
\newcommand{\CQG}{Class. Quantum Grav.}
\newcommand{\parallelsum}{\mathbin{\!/\mkern-5mu/\!}}

\title{The Steep Price of No Hair in a Modified Loop Quantum Cosmology}
\author{Meysam Motaharfar$^1$}
\email{mmotah4@lsu.edu}
 \author{Parampreet Singh$^{1,2}$}
\email{psingh@lsu.edu}
\affiliation{$^1$ Department of Physics and Astronomy, 
Louisiana State University, Baton Rouge, LA 70803, USA}
 \affiliation{$^2$ Center for Computation and Technology,
Louisiana State University, Baton Rouge, LA 70803, USA}

\begin{abstract}
\textcolor{black}{
A modified version of loop quantum cosmology model, the so-called mLQC-I, motivated by Thiemann’s regularization of the Hamiltonian constraint, leads to the resolution of the big bang singularity and a bounce in the isotropic setting, where either the pre-bounce or post-bounce epoch is necessarily characterized by an emergent Planckian de Sitter phase. In this work we explore the Planckian physics of this mLQC-I prescription for the Bianchi-I spacetimes. We show that, as in the isotropic model, there exists an emergent de Sitter phase which naturally dampens anisotropic shear and removes cosmic hair. However, this isotropization comes at a steep price: although a macroscopic post-bounce regime is achieved, the universe does not become classical. As is well known from the Kasner solution, the classical evolution of a contracting Bianchi-I universe toward the singularity can in general be either point-like or cigar-like. However, cigar-like evolution is prevalent unless the matter content dominates over the anisotropic shear. For a class of physically admissible initial conditions corresponding to cigar-like evolution, we further demonstrate that this isotropization mechanism is non-generic. These results clarify and reinterpret recent claims by Gan et al. \cite{Gan:2025uvt} that, in anisotropic mLQC-I, quantum gravity effects generically damp anisotropic shear in a manner independent of initial conditions and matter content, and that this damping arises from a novel quantum gravity effect. Our work explains the origin of this mechanism and its limitations in the mLQC-I model.}

\end{abstract}

\maketitle

\section{Introduction} 

Cosmological observations confirm that the universe, as observed today, is expanding, spatially flat, homogeneous, and isotropic on large scales \cite{SupernovaSearchTeam:1998fmf, Planck:2018vyg}. While the standard big bang cosmology can elegantly explain most of the expansion history of the universe, it faces several problems, including the horizon and flatness problems, which together imply unnaturally fine-tuned initial conditions for the universe. To address these problems, one can introduce various mechanisms to flatten, homogenize, and isotropize the universe, starting from rather general initial conditions, including an anisotropic and inhomogeneous universe with arbitrary spatial geometry. Among several proposals, inflationary and ekpyrotic cosmologies have been well-studied and have proven to be successful in achieving this goal to different extents \cite{Guth:1980zm, Linde:1981mu, Khoury:2001wf}. In fact, cosmic no-hair theorems, first proved in Ref. \cite{Wald:1983ky}, demonstrate that under certain conditions, the late-time behavior of any accelerating anisotropic universe asymptotically evolves towards an isotropic universe (also see \cite{Erickson:2003zm} if one assumes ekpyrosis). Despite their tremendous success, these models of the early universe are incomplete, as they do not provide
a resolution of the big bang singularity, where the spacetime curvature diverges, signifying the breakdown of classical general relativity.
This implies that a theory of quantum gravity is necessary to resolve the big bang singularity and make physical predictions at the Planck scale. An interesting question is whether the new physics at the Planck scale can isotropize the universe independent of the matter content, fine-tuning, or exotic ingredients. In fact, it would be theoretically and conceptually appealing if the same quantum gravitational effects responsible for the resolution of the big bang singularity, also lead into a generic dynamical isotropization mechanism for the universe. It is then pertinent to investigate this question in theories of quantum gravity in which quantum gravitational effects generically resolve the big bang singularity. The goal of this manuscript is, therefore, to study the viability of a potential quantum gravitational isotropization mechanism within the loop quantum cosmology (LQC) framework, in particular in the approach of a modified loop quantum cosmology (mLQC) motivated from Thiemann's regularization of Hamiltonian constraint.

LQC applies techniques from loop quantum gravity, a non-perturbative and background-independent canonical quantization of gravity, to symmetry reduced cosmological spacetimes \cite{Ashtekar:2011ni}. A key prediction of this framework is that the big bang singularity is replaced with a quantum bounce in the Planck regime \cite{Ashtekar:2006rx,  Ashtekar:2006wn}. For a spatially flat, homogeneous, and isotropic universe with a massless scalar field playing the role of clock, which turns out to be an exactly solvable model \cite{Ashtekar:2007em}, many contrasting results in comparison to Wheeler-DeWitt theory have been obtained. For example, it can be shown that the bounce occurs for all the states in the physical Hilbert space \cite{Ashtekar:2007em}; the probability of singularity resolution is exactly unity in an exactly solvable model of LQC \cite{Craig:2013mga}; and for an initialized state in either the contracting or expanding branch, the Krylov complexity remains finite at the bounce  \cite{Motaharfar:2025gxz}. At the fundamental level, singularity resolution in LQC arises due to underlying quantum geometry, which causes the spacetime curvature to be bounded. The resolution of big bang singularity has also been generalized to other spacetimes, including in the presence of spatial curvature \cite{close-open-lqc}, inflationary potential \cite{Giesel:2020raf}, anisotropies \cite{Ashtekar:2009vc, Ashtekar:2009um, Wilson-Ewing:2010lkm,Martin-Benito:2009xaf,Diener:2017lde}, and in the presence of Fock quantized inhomogeneities \cite{Garay:2010sk}. Interestingly, extensive numerical simulations have demonstrated that the underlying quantum evolution can be captured using an effective spacetime description \cite{Diener:2014mia, Diener:2017lde}. Assuming the validity of effective dynamics, the phenomenological implications of LQC for several cosmological models have been extensively studied (see Refs. \cite{Li:2023dwy, Li:2021mop} and references therein for a comprehensive review). Importantly, the bounds on the energy density, expansion, and anisotropic shear scalar in different models have been found \cite{Gupt:2011jh, Singh:2013ava, McNamara:2022dmf}, and strong curvature singularities have been shown to be generically resolved in isotropic models \cite{Singh:2009mz,  Saini:2018tto} as well as anisotropic models \cite{strong-singularity-resolution}. It has also been shown that  a viable inflationary model can be constructed from highly
anisotropic initial conditions \cite{Gupt:2013swa, Martineau:2017sti}. 

However, as in any theory of quantum gravity, there are quantization and regularization choices in LQC. In the standard LQC model, the Lorentzian and Euclidean terms in the Hamiltonian constraint are combined using classical symmetry before quantization in Friedmann-Lemaître-Robertson-Walker (FLRW) spacetime. However, if these two terms are treated independently following the construction of loop quantum gravity, they lead to different inequivalent quantizations of LQC arising from Thiemann’s regularization of the Hamiltonian constraint. In doing so, one can use classical identities on gravitational phase space to express the extrinsic curvature of the Lorentzian term of the Hamiltonian constraint in terms of holonomies \cite{Thiemann}. The result is different versions of modified LQC \cite{Yang:2009fp, dapor-liegener, Li:2018opr}. Of these, the most prominent one is   mLQC-I which has some intriguing features \cite{Yang:2009fp, dapor-liegener, Assanioussi:2018hee, Li:2018opr, Li:2018fco}, and can also leave distinct potential observational effects in the CMB \cite{mlqc-I} in comparison to standard LQC. For the isotropic model, the quantum Hamiltonian constraint yields a fourth-order quantum difference equation rather than a second-order equation in standard LQC  \cite{Saini:2019tem}. It turns out that singularities are generically resolved as in the standard LQC \cite{Saini:2018tto}. Interestingly, it has been shown that mLQC-I leads into an emergent de Sitter spacetime that arises with a Planckian value in either the pre-bounce or post-bounce branch depending on the initial conditions \cite{Li:2018opr,Assanioussi:2018hee}. This implies that the de Sitter branch is purely of a quantum gravitational origin. Due to the emergence of the Planck scale cosmological constant in either the pre-bounce or the post-bounce regime, it can also be shown that mLQC-I does not allow a cyclic evolution \cite{Li:2021fmu}. To obtain a viable universe such as ours, the only choice in mLQC-I is to have initial conditions such that the Planckian cosmological constant is in the pre-bounce branch \cite{Li:2018opr}. The Thiemann regularized prescription of mLQC-I has also been generalized to anisotropic Bianchi-I spacetimes \cite{Garcia-Quismondo:2019kav, Garcia-Quismondo:2019dwa}. However, given the complicated form of the Hamiltonian and the resulting Hamilton's equations, various physical implications remained unexplored so far. 

Recently, some implications of quantum gravitational effects on anisotropic shear have been studied in mLQC-I generalized to Bianchi-I spacetimes, and it is claimed that gravitational effects lead to a generic dynamical isotropization mechanism \cite{Gan:2025uvt}. Considering initial conditions corresponding to a universe that is large, classical, and contracting, in the pre-bounce branch, the time evolution of directional scale factors for a universe filled with different matter content such as dust, radiation, massless scalar field, the inflaton field, and the ekpyrotic field is computed. It is found that even though three directional scale factors have different expansion rates in the pre-bounce contracting branch, they rapidly expand to an isotropic universe in the post-bounce branch. The anisotropic shear is found to be dynamically suppressed, which decays rapidly to zero within the deep quantum regime for all initial conditions and matter content. The analysis claims that quantum gravitational effects are able to isotropize the universe in the post-bounce expanding branch without any need for adding exotic ingredients for isotropization, such as an ekpyrotic or inflation field. If this turns out to be the case, it can result in a novel way to obtain an isotropized classical universe using loop quantum gravity effects. However, an important question that was not investigated so far is whether the macroscopic universe obtained after isotropization becomes classical. At first it may seem that the existence of a macroscopic universe implies classicality. However, earlier studies show that this may not always be the case (see, e.g., \cite{Motaharfar:2023hil}). As mentioned earlier, in mLQC-I there is an emergence of the Planckian cosmological constant in the isotropic setting in which the universe retains its quantum character and does not become classical. A natural question is whether such a phase also exists in the anisotropic setting, which can cause the loss of all cosmic hair by damping the anisotropic shear. Further, in isotropic mLQC-I, the initial conditions must be chosen such that this de Sitter phase is in the pre-bounce branch. Otherwise, the universe remains in the quantum regime in the post-bounce branch forever. Since the anisotropic mLQC-I is a generalization of the corresponding  isotropic model, understanding the issue of classicality becomes important in the anisotropic setting, which can shed useful insights on the underlying mechanism of the anisotropic shear damping and whether it is phenomenologically viable.

To clarify these issues, in this manuscript, we revisit the effective dynamics of mLQC-I in the context of Bianchi-I spacetimes to investigate the implications of quantum gravitational effects on the dynamics of anisotropy across the quantum bounce. In particular, we focus on understanding the details of isotropization of the universe from anisotropic initial conditions to gain insights on its quantum gravitational origin. \textcolor{black}{A key part of this analysis is the vacuum Bianchi-I spacetime. The vacuum sector is important in its own right because it provides the cleanest setting to isolate the interplay between anisotropic shear, quantum geometry, and classicality, without additional matter effects. It also serves as a useful test of whether isotropization is truly generic in the effective dynamics of mLQC-I. Our analysis is based on the effective spacetime dynamics of Bianchi-I mLQC-I. In this framework, once the lapse is fixed, the effective Hamilton equations define a well-posed dynamical system in cosmic time, and our conclusions about classicality, isotropization, and genericity are made within this effective description.}

We find that quantum gravitational isotropization of the universe as found earlier for the Bianchi-I version of mLQC-I does indeed occur, but it is rather problematic as well as non-generic. Our work clarifies the so far unknown mechanism underlying this effect.  We show that it is tied to the emergence of de Sitter spacetime with Planckian value in the post-bounce branch, similar to isotropic mLQC-I. Since the isotropic model is a limit of the Bianchi-I model when anisotropies vanish, it is natural to expect that certain unique features of mLQC-I, which distinguish it from standard LQC, are captured in the anisotropic setting. In particular, we find that while the scale factor of the universe is macroscopic in the post-bounce regime, the spacetime curvature remains Planckian. The universe after the quantum bounce remains in the Planckian regime in the post-bounce branch with no exit to the classical universe. The emergence of a de Sitter universe with a macroscopic scale factor and Planckian curvature exactly mirrors the situation in the isotropic model for mLQC-I.   We  further show that isotropization does not occur for Bianchi-I vacuum spacetime, nor for a range of initial conditions when the universe is sourced with a perfect fluid. Taken together, these results indicate that the quantum gravitational isotropization mechanism observed in anisotropic mLQC-I comes at a substantial cost of classicality. And, even after paying such a steep price, the mechanism is not generic. 

\textcolor{black}{
Ref. \cite{Gan:2026ivx} recently questioned the physical admissibility of the initial conditions used in our simulations. However, from the classical Kasner behavior it is well known that, as a contracting Bianchi-I universe approaches the big bang singularity, the evolution can in general be either point-like, with all three scale factors contracting, or cigar-like, with two scale factors contracting and one expanding. Which of these behaviors is realized depends on the interplay between matter content and anisotropic shear. In particular, when matter does not dominate over the anisotropic shear, the approach to the singularity is generically cigar-like. The initial conditions considered in this work are therefore physically admissible and, in this regime, more prevalent than the point-like data emphasized in \cite{Gan:2026ivx}. This strengthens our conclusion that isotropization in mLQC-I is non-generic: rather than being a universal feature, it occurs only for a finely tuned subclass of point-like evolutions considered in Ref. \cite{Gan:2026ivx}.}

The manuscript is structured as follows. We briefly review the dynamics of the classical Bianchi-I spacetimes in terms of Ashtekar-Barbero variables and discuss the solution for a universe filled with a cosmological constant in Section \ref{Section II}. We then briefly present the effective dynamical equations for a modified LQC -- in particular mLQC-I, applied to Bianchi-I spacetimes in Section \ref{Section III}. We next provide numerical analysis for the evolution of the universe considering the vacuum spacetime and also for a universe filled with a perfect fluid with a barotropic equation of state, such as dust, radiation, and a massless scalar field, in Section \ref{Section IV}. We discuss the cases when isotropization occurs and also when it is absent. Note that in this section, we use Planck units. Finally, we summarize the main results in Section \ref{Section V}. It is important to note that not all variants of LQC based on Thiemann regularization of the Hamiltonian constraint result in an emergent de Sitter phase. As an example, there exists mLQC-II which has similar bounce as in standard LQC and for which classical limit exists on both sides of the bounce \cite{Li:2018fco}. The results of this manuscript would not apply to such modified versions of LQC. Our investigation pertains only to anisotropic version of mLQC-I.

\section{Classical Bianchi-I Spacetimes with a Cosmological Constant}\label{Section II}

Since we later show that the underlying physics behind the quantum gravitational isotropization mechanism is tied to the emergence of a Planckian cosmological constant, in this section we briefly summarize the classical dynamics of Bianchi-I spacetimes with a cosmological constant. The loop quantization program is based on the classical gravitation phase space variables written in terms of the Ashtekar-Barbero connection $A^{i}_{a}$ and the triads $E^{a}_{i}$ (where $i = 1, 2, 3$); therefore, it is useful to understand the dynamics of Bianchi-I spacetimes in the canonical variables and how they are related to the conventional metric variables. To this end, we consider a spatially flat homogeneous (but anisotropic) universe described by Bianchi-I spacetimes 
\begin{align}\label{Bianchi-I metric}
\mathrm{d}s^2 = - N^2(t)\, \mathrm{d}t^2 + a_{1}^2(t) \, \mathrm{d}x^2 + a_{2}^2(t)\, \mathrm{d}y^2 + a_{3}^2(t) \, \mathrm{d}z^2,
\end{align}
where $N(t)$ is the lapse function and $a_{i}(t)$ are directional scale factors, while the mean scale factor is defined as $a \coloneqq (a_{1}a_{2}a_{3})^{1/3}$. In the isotropic limit, all the scale factors are equal, i.e., $a_{1} = a_{2} = a_{3} = a$, and the metric (\ref{Bianchi-I metric}) reduces to the Friedmann-Lemaître-Robertson-Walker (FLRW) metric describing a spatially flat, homogeneous, and isotropic universe. 

The Ashtekar-Barbero variables $A_{a}^{i}$ and $E_{i}^{a}$ reduce to connections $c_{i}$ and triads $p_{i}$ with only one independent component per spatial direction upon symmetry reduction and imposing the Gauss and the spatial-diffeomorphism constraints. The triads are kinematically related to the directional scale factors, i.e., $|p_{i}| = a_{j}a_{k}$ ($i \neq j \neq k$), where the modulus arises due to the orientation of the triads. We assume the orientations of the triads to be positive. Hence, the classical Hamiltonian constraint for matter content minimally coupled to gravity in terms of $c_{i}$ and $p_{i}$ is given by
\begin{align}\label{Hamiltonian}
\mathcal{H}_{\textrm{cl}} = \mathcal{H}_{G} + \mathcal{H}_{M} = - \frac{N}{8 \pi G \gamma^2 v} \left(c_{1}p_{1} c_{2}p_{2} + c_{2}p_{2}c_{3}p_{3} + c_{3}p_{3}c_{1}p_{1}\right) + \mathcal{H}_{M},
\end{align}
where $\mathcal{H}_{M}$ is the matter part of the Hamiltonian, $v = \sqrt{p_{1}p_{2}p_{3}}$ is the physical volume of a unit comoving cell\footnote{In non-compact models in LQC, one introduces a fiducial cell to define a symplectic structure whose coordinate lengths enter the relation between triads and scale factors. We fix the volume of this fiducial cell to be unity.}, and $\gamma \approx 0.2375$ is the Barbero-Immirzi parameter, which is fixed by black hole thermodynamics in loop quantum gravity. Given the classical Hamiltonian constraint, one can find Hamilton's equation as follows:
\begin{align} \label{Hamilton's equations}
\dot p_{i} = \{p_{i}, \mathcal{H}_{\textrm{cl}}\} = - 8 \pi G \gamma \frac{\partial \mathcal{H}_{\textrm{cl}}}{\partial c_{i}}, \ \ \ \ \ \ \ \ \ \ \ \ \dot c_{i} = \{c_{i}, \mathcal{H}_{\textrm{cl}}\} = 8 \pi G \gamma \frac{\partial \mathcal{H}_{\textrm{cl}}}{\partial p_{i}},
\end{align}
where we have set the lapse function to unity, i.e., $N = 1$. Hence, a dot denotes the derivative with respect to cosmic time $t$. A set of equations for $p_{i}$ reveals that in the classical regime, $c_{i} = \gamma \dot a_{i}$. One can use the Hamiltonian constraint (\ref{Hamiltonian}) to find the evolution equations in terms of directional Hubble rates $H_i = \dot a_i/a_i$: 
\begin{align}
H_{1}H_{2}+ H_{2}H_{3} + H_{3}H_{1} &= 8 \pi G \rho, \label{Friedmann-I}\\
{\color{black}{H_{2} H_3}} + H_{2}^2 + H_{3}^2 + \dot H_{2} + \dot H_{3} &= - 8 \pi G P, \label{Friedmann-II}
\end{align}
\textcolor{black}{(\textrm{and its cyclic permutations}). Here,} we considered a perfect fluid with a barotropic equation of state, i.e., $\omega = P/\rho$, while $\rho = \mathcal{H}_{M}/v$ and $P = - \frac{\partial \mathcal{H}_{M}}{\partial v}$. In terms of the triads, $H_{i} = \dot a_{1}/a_{1} = (\dot p_{j}/p_{j} + \dot p_{k}/p_{k} - \dot p_{i}/p_{i})/2$. Further, the mean Hubble parameter is defined as $H = \sum_{i}H_{i}/3$. One can easily check that in the isotropic limit, Eqs. (\ref{Friedmann-I}) and (\ref{Friedmann-II}) reduce to the Friedmann equations for a spatially flat, homogeneous, and isotropic spacetime: 
\begin{align}\label{Friedmann-i}
H^2 = \frac{8\pi G}{3} \rho, \ \ \ \ \ \ \ \ \ \ \ \ \ \ \ \ \ \dot H = - 4\pi G (\rho + P).
\end{align}
One can also rewrite Eqs. (\ref{Friedmann-I}) and (\ref{Friedmann-II}) as generalized isotropic Friedmann equations with the anisotropic shear scalar as follows:  
\begin{align}\label{generalized}
H^2 = \frac{8 \pi G}{3} \rho + \frac{\sigma^2}{6}, \ \ \ \ \ \ \ \ \ \ \ \ \ \ \dot H = - 4\pi G \left(\rho + P\right) - \frac{ \sigma^2}{2} .
\end{align}
Here the anisotropic shear scalar $\sigma^2$, which measures the deviation from isotropic spacetime, is a traceless part of the expansion tensor, which in terms of directional Hubble parameters $H_{i}$ is given by
\begin{align}\label{shear-tensor}
\sigma^2 = \sigma^{\mu\nu} \sigma_{\mu\nu} = \frac{1}{3} \left(\left(H_{1}- H_{2}\right)^2 + \left(H_{2}- H_{3}\right)^2 + \left(H_{3}- H_{1}\right)^2 \right) .
\end{align}
It is related to the mean scale factor as $\sigma^2 = \Sigma^2/a^6$ with $\Sigma$ being a constant. This implies that the anisotropic shear $\Sigma^2$ is a constant of motion in the classical Bianchi-I spacetimes, assuming isotropic matter content. Note that in the isotropic limit $\sigma^2=0$ and Eqs. (\ref{generalized}) reduce to the Friedmann equations for a spatially flat, homogeneous, and isotropic universe given by Eqs. (\ref{Friedmann-i}). 

Given the dynamical equations (\ref{Friedmann-I}) and (\ref{Friedmann-II}), it is straightforward to establish that a classical Bianchi-I universe will isotropize in the presence of a positive cosmological constant. Let us consider a Bianchi-I universe filled with a positive cosmological constant $\Lambda$, i.e., $8\pi G \rho = \Lambda$. One can find the following well-known Kasner solution \cite{Stephani:2003tm}\footnote{While the classical Kasner solutions are singular, their non-singular variants have been extensively studied in LQC (see, e.g., \cite{Gupt:2012vi, Singh:2016jsa, Wilson-Ewing:2017vju, deCesare:2019suk}).}
\begin{align}
    a_{i}(t) = \lambda^{-k_{i}}\cosh^{\frac{2}{3}}(\lambda t) \tanh^{k_{i}}(\lambda t), 
\end{align}
where $\lambda = \sqrt{3\Lambda}/2$ and $k_{i}$ are the Kasner exponents. The Kasner exponents are constant and satisfy $k_{1}+k_{2}+k_{3} =1$ and $k_{1}^2+k_{2}^2 + k_{3}^2=1$. In the late-time limit, i.e., $t\rightarrow +\infty$, all scale factors behave as $a(t) \propto e^{\sqrt{\Lambda/3}t}$. This corresponds to the well-known solution to a de Sitter spacetime. The energy density of the cosmological constant remains constant while the anisotropic shear $\sigma^2$ decays rapidly. As a result, the generalized Friedmann equation (\ref{generalized}) can be approximated by Eqs. (\ref{Friedmann-i}). Thus, there is an isotropization of the Bianchi-I universe in the presence of a positive cosmological constant. Note that at late times, the curvature invariants such as the Ricci scalar and Kretschmann scalar will be proportional to the cosmological constant, i.e., $R = 4\Lambda$ and $K = R_{abcd} R^{abcd} = 8\Lambda^2/3$ as in the de Sitter universe. We will see a similar behavior in the loop quantum evolution of the Bianchi-I spacetimes in mLQC-I even without the cosmological constant, which will confirm the emergent de Sitter behavior from quantum geometry.

\section{Effective dynamical equations of mLQC-I in Bianchi-I spacetimes}\label{Section III}

In loop quantum gravity, the gravitational part of the Hamiltonian constraint consists of two parts: the Euclidean part $\mathcal{H}_{E}$ and the Lorentzian part $\mathcal{H}_{L}$. Then, one needs to regularize these two terms independently using Thiemann's regularization techniques \cite{Thiemann} to quantize the classical Hamiltonian constraint. However, in the standard LQC, before quantization, the Euclidean and Lorentzian terms are combined by using their proportionality at the classical level. The same procedure is followed in the quantization of the Bianchi-I spacetimes. In general, one could treat the Euclidean and Lorentzian parts independently in the quantization procedure. It turns out that this procedure leads to an inequivalent quantization of the same classical spacetime. In particular, using the classical identities on the gravitational phase space, one can express the extrinsic curvature of the Lorentzian term of the gravitational Hamiltonian constraint in terms of holonomies, leading into the so-called Thiemann regularized loop quantum cosmologies. Of these a prominent one is modified loop quantum cosmology-I (mLQC-I) for isotropic spacetimes \cite{Yang:2009fp, dapor-liegener, Assanioussi:2018hee, Li:2019ipm, Li:2018opr, Li:2021mop}. Recently mLQC-I has been generalized to anisotropic Bianchi-I spacetimes\cite{Garcia-Quismondo:2019kav, Garcia-Quismondo:2019dwa}, using Thiemann's regularization procedure for the Euclidean and Lorentzian parts as follows:
\begin{align}\label{effective Hamiltonian}
\mathcal{H}^{\textrm{eff}}_{G} = \mathcal{H}_{E} + \mathcal{H}_{L} \approx 0, 
\end{align}
where $\mathcal{H}_{E}$ and $\mathcal{H}_{L}$ are given by 
\begin{align}
\mathcal{H}_{E} = \frac{1}{8 \pi G v} \left(\frac{p_{1}\sin (\bar \mu_{1} c_{1})}{\bar \mu_{1}}\frac{p_{2}\sin (\bar \mu_{2} c_{2})}{\bar \mu_{2}}  + \textrm{cyclic permutations}\right),
\end{align}
and
\begin{align}
\nonumber \mathcal{H}_{L} & = -\frac{1}{8 \pi G v} \frac{1+\gamma^2}{4\gamma^2} \Bigg[\left(\frac{ p_{1}\sin (\bar \mu_{1} c_{1})}{\bar \mu_{1}} \frac{p_{2}\sin (\bar \mu_{2} c_{2})}{\bar \mu_{2}}\right) \left(\cos(\bar \mu_{1}c_{1}) + \cos(\bar \mu_{3}c_{3}))(\cos(\bar \mu_{2}c_{2}) + \cos(\bar \mu_{3}c_{3})\right)  \\& ~~~ + ~~~ \, \textrm{cyclic permutations} \Bigg] . 
\end{align}
Here $\bar \mu_{i}$ are the functions of the triads, i.e., $\bar \mu_{i} = \sqrt{\Delta p_{i}/(p_{j}p_{k})}$ with $\Delta = 4\pi \sqrt{3}\gamma l_{\textrm{Pl}}^2$ being the minimum eigenvalue of the area operator in loop quantum gravity, and $l_{\textrm{Pl}}$ is the Planck length. The effective gravitational Hamiltonian constraint (\ref{effective Hamiltonian}) reduces to its classical counterpart in the limit $\mu_{i}c_i \rightarrow 0$ (or more accurately, when $\mu_{i}c_i \rightarrow n \pi$ with $n$ being an integer). In this limit, the $\sin (\bar \mu_{i}c_{i})$ terms in the effective gravitational Hamiltonian constraint reduce to the classical connection variables, $c_{i}$, specifically
\begin{align}\label{classicality condition}
    \lim_{\bar \mu_{i} c_{i}\rightarrow n\pi} \left(\frac{\sin (\bar \mu_{i}c_{i})}{\bar \mu_{i}}\right) \approx c_{i}.
\end{align}
Hereafter, we refer to this condition as the “classicality condition.” Moreover, the effective Hamiltonian constraint (\ref{effective Hamiltonian}) reduces to the effective Hamiltonian constraint for the isotropic mLQC-I when $\bar \mu_{1} = \bar \mu_{2} = \bar \mu_{3} = \sqrt{\Delta/p}$, where $p = p_{1} = p_{2} = p_{3}$. 

Given the effective Hamiltonian constraint (\ref{effective Hamiltonian}), one can then find the effective Hamilton's equations  for the triads $p_{i}$ and connections $c_{i}$ using:
\begin{align} \label{Hamilton's equations}
\dot p_{i} = \{p_{i}, \mathcal{H}_{\textrm{eff}}\} = - 8 \pi G \gamma \frac{\partial \mathcal{H}_{\textrm{eff}}}{\partial c_{i}}, \ \ \ \ \ \ \ \ \ \ \ \ \dot c_{i} = \{c_{i}, \mathcal{H}_{\textrm{eff}}\} = 8 \pi G \gamma \frac{\partial \mathcal{H}_{\textrm{eff}}}{\partial p_{i}} .
\end{align}
{\color{black} Here $\mathcal{H_{\textrm{eff}}} = \mathcal{H}_{G}^{\textrm{eff}} + \mathcal{H}_{M}$ and the effective Hamilton's equations for $p_{i}$ are given by \cite{Gan:2025uvt}

\begin{equation}
\label{pi}
\frac{\dot{p}_i}{p_i} = - \frac{\gamma}{\sqrt{\Delta}}
\left({\text{sn}}_i 
f^{\text{A}}_{ijk} +
{\text{sn}}_j 
f^\text{B}_{i,ik,jk} 
+
{\text{sn}}_k 
f^\text{B}_{i,ij,jk}
\right) ,
\end{equation}
where $(i,j,k) ~ \text{is a permutation of} ~ (1,2,3)$. We have used the short-hand notation ${\text{cs}}_i \equiv \cos(\bar{\mu}_i c_i)$, and ${\text{sn}}_{i} \equiv \sin(\bar{\mu}_i c_i)$, and the functions $f^\text{A}_{ijk}$ and $f^\text{B}_{i,jk,lm}$ are defined as follows \cite{Gan:2025uvt}:
\begin{align}
f^\text{A}_{ijk} 
& \equiv 
\delta_{\gamma}[{\text{sn}}_i({\text{sn}}_j + {\text{sn}}_k)({\text{cs}}_j + {\text{cs}}_k)  + {\text{sn}}_j{\text{sn}}_k(2{\text{cs}}_i + {\text{cs}}_j + {\text{cs}}_k)]\\
f^\text{B}_{i,jk,lm} 
&\equiv 
{\text{cs}}_i[1 - \delta_{\gamma}({\text{cs}}_j + {\text{cs}}_k)({\text{cs}}_l + {\text{cs}}_m)].
\end{align}
with $\delta_{\gamma} \equiv (1+\gamma^2)/(4\gamma^2)$. Similarly, effective Hamilton's equations for $c_{i}$ can be written as follows:
\begin{eqnarray}\label{ci}
\dot{c}_i  &=& \frac{\gamma}{2p_i v} \big[
{c_i p_i} \,  
\left(p_j \,\mu\mathrm{sn}_j  f^\text{B}_{i,ik,jk} 
+ p_k \, \mu\mathrm{sn}_k  f^\text{B}_{i,ij,jk}
\right) \nonumber\\
&& - c_j p_j \,  
\left(p_i \, \mu\mathrm{sn}_i  
f^\text{B}_{j,ik,jk}
+  p_k \, \mu\mathrm{sn}_k  
f^\text{B}_{j,ij,ik}
\right) \nonumber\\
&& -
c_k p_k \,  
\left(p_i \, \mu\mathrm{sn}_i 
f^\text{B}_{k,ij,jk}
+  p_j \, \mu\mathrm{sn}_j  
f^\text{B}_{k,ij,ik}
\right) \nonumber\\
&& + p_j \, \mu\mathrm{sn}_j  p_k \, \mu\mathrm{sn}_k  
g^{-}_{ij,ik}
+
 p_i \, \mu\mathrm{sn}_i  p_k \, \mu\mathrm{sn}_k  
g^{+}_{ij,jk} \nonumber\\
&& +
 p_i \, \mu\mathrm{sn}_i  p_j \, \mu\mathrm{sn}_j  
g^{+}_{ik,jk}
\big] \nb\\
&& + 
\frac{4\pi G \gamma v}{p_i} \left( \rho_M + 2p_i \frac{\partial \rho_M}{\partial p_i} \right),
\end{eqnarray}
where $g^{\pm}_{ij,lk}$
\begin{align}
g^{\pm}_{ij,lk}
& \equiv 
1 
- 
\delta_\gamma [(\mathrm{cs}_i + \mathrm{cs}_j)(\mathrm{cs}_l + \mathrm{cs}_k)
\pm
(\mathrm{cs}_i + \mathrm{cs}_j)
(
\bar{\mu}_l c_l \, \mathrm{sn}_l
\pm 
\bar{\mu}_k c_k \, \mathrm{sn}_k 
)
\\ &-(\mathrm{cs}_l + \mathrm{cs}_k)
(
\bar{\mu}_i c_i \, \mathrm{sn}_i
- 
\bar{\mu}_j c_j \, \mathrm{sn}_j 
)
],
\end{align}
with $\mu sn_{i} \equiv sn_{i}/\bar\mu_{i}$.} For the classicality condition (\ref{classicality condition}), one expects these effective equations to reduce to the classical ones, recovering the dynamics of classical Bianchi-I spacetimes. In other words, for a consistent picture in LQC, a large contracting classical universe with small spacetime curvature should bounce to a large expanding classical universe with small spacetime curvature. In addition, it is to be expected that in the isotropic limit, the above equations yield the effective Hamilton's equations for mLQC-I. Since the latter has an emergent de Sitter phase, one expects the above equations to demonstrate a similar behavior at least for some choice of matter fields and initial conditions.

Before we close this section, it is important to note that the emergence of a macroscopic universe from effective Hamilton's equations does not imply classicality. In the case of anisotropic models, this was demonstrated for the first time in Ref. \cite{Motaharfar:2023hil}, where it was shown that even if one can fix the initial conditions for the triads and connections in the pre-bounce branch (both contracting and expanding universes) to have a large macroscopic classical universe, it is not guaranteed that the universe becomes classical as it passes through the bounce. This result was used to constrain quantization ambiguities in the loop quantization of Bianchi-I spacetimes. Since the effective Hamilton's equations are very complicated, it is not obvious that classical universes in the pre-bounce branch become classical in the post-bounce branch. Hence, it is important to numerically solve the effective Hamilton's equation and carefully analyze whether the universe becomes classical in the post-bounce branch for the considered initial conditions. 

\section{Numerical Analysis of Effective Dynamical Equations}\label{Section IV}

Given the effective Hamiltonian constraint and the corresponding Hamilton's equations (\ref{pi}) and (\ref{ci}), we are interested in finding the evolution of the universe across the bounce for the vacuum spacetime and also for a universe filled with different matter content, namely radiation, dust, and a massless scalar field. We fix the initial value of triads and connections such that the universe starts in the contracting branch. We consider the initial values of two connections to be negative and then fix the initial value of the third connection using the effective Hamiltonian constraint. To ensure that we are in the classical regime, we tune the initial conditions such that all three $\bar\mu_{i}c_{i}$ are close to $n\pi$. Finally, we numerically solve effective Hamilton's equations and find the time evolution of physical quantities such as directional scale factors, mean scale factor, anisotropic shear scalar, and curvature invariants and also carefully explore the classicality conditions. Interestingly, our results demonstrate that the quantum gravitational isotropization mechanism observed in generalization of mLQC-I to Bianchi-I spacetimes occurs due to an emergence of a Planckian de Sitter universe in the post-bounce branch. The emergence of this Planckian cosmological constant in the post-bounce branch is expected, as it also appears in the isotropic mLQC-I, which is the isotropic limit of mLQC-I in anisotropic setting when the anisotropic shear scalar vanishes. We show that such an isotropization mechanism, though interesting, suffers from two problems: first, the post-bounce universe does not become classical even in the macroscopic limit due to the Planckian value of the emergent cosmological constant; second, the mechanism is not generic, as it does not occur for a range of initial conditions and for some types of matter considered in this manuscript.

\subsection{Emergent Planckian de Sitter Universe and Isotropization}

\begin{figure}
    \centering
    \includegraphics[width=0.47\linewidth]{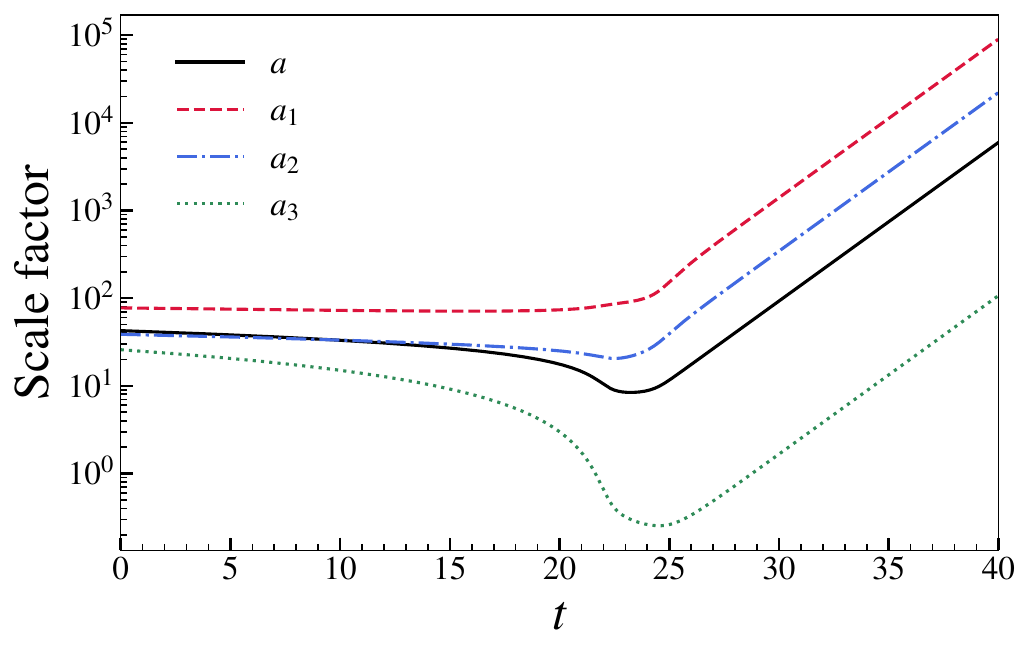}
    \caption{The time evolution of the directional scale factors $(a_i)$ and the mean scale factor $(a)$ is shown for the initial conditions $p_{1}=1000$, $p_{2}=2000$, $p_{3}=3000$, $c_{1}=-0.13$, $c_{2}=-0.12$, and dust field energy density $\rho_{m}=3.55 \times 10^{-5}$. There is an emergent Planckian cosmological constant in the post-bounce branch, as a result of which the universe becomes isotropic, behaving like a de Sitter universe in the late-time. Hereafter, all physical quantities are evaluated in Planck units.}
\label{dust}
\end{figure}

\begin{figure}
    \centering
    \includegraphics[width=0.43\linewidth]{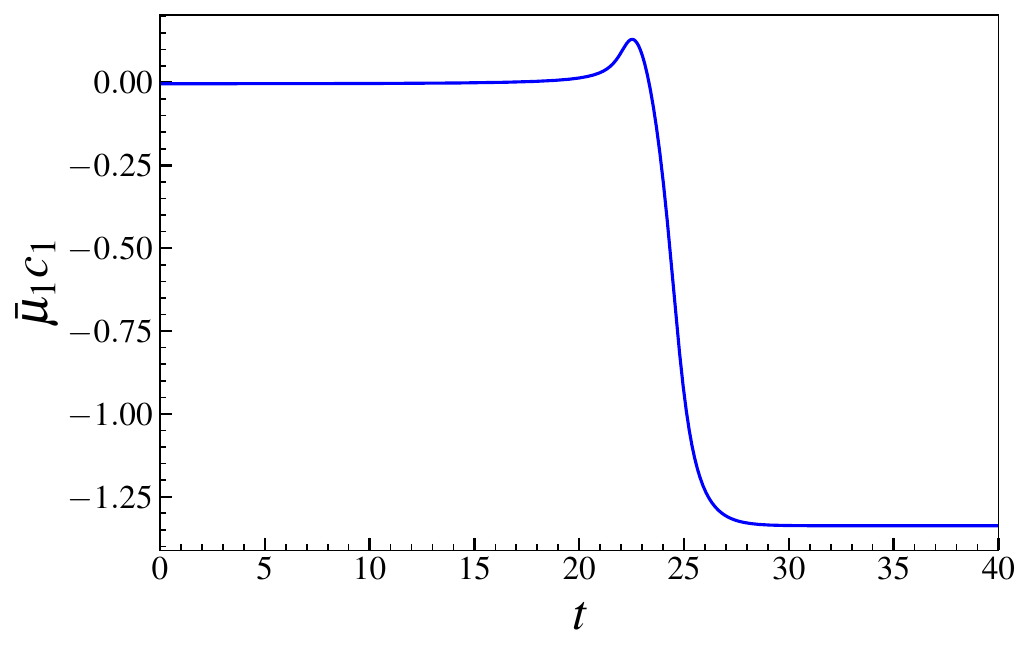} \ \ \ \ \ \ \ \ \ \ 
    \includegraphics[width=0.43\linewidth]{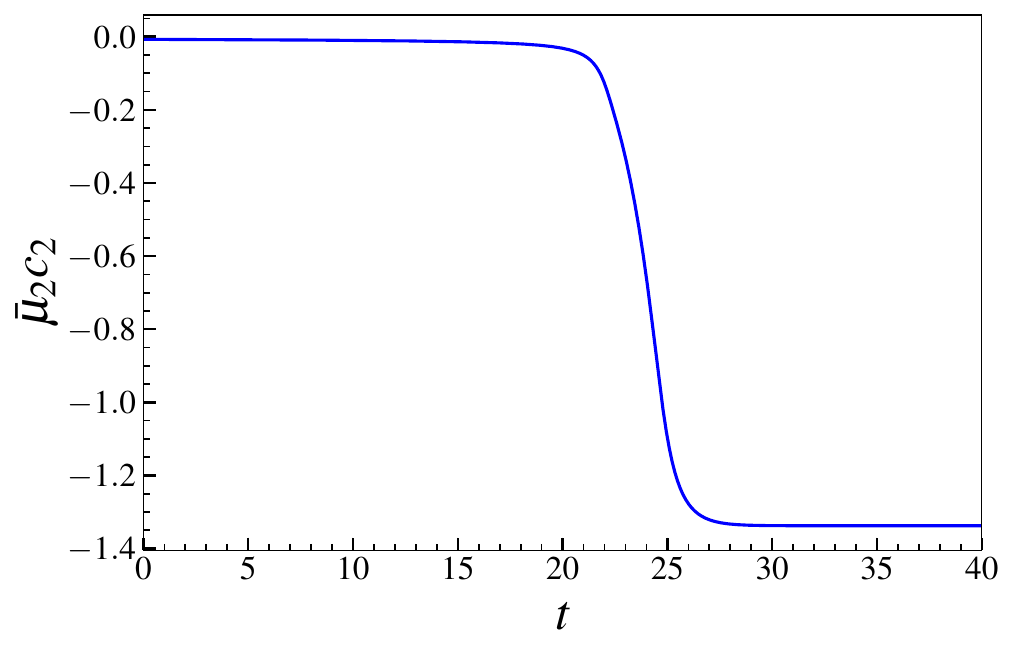}
    \includegraphics[width=0.43\linewidth]{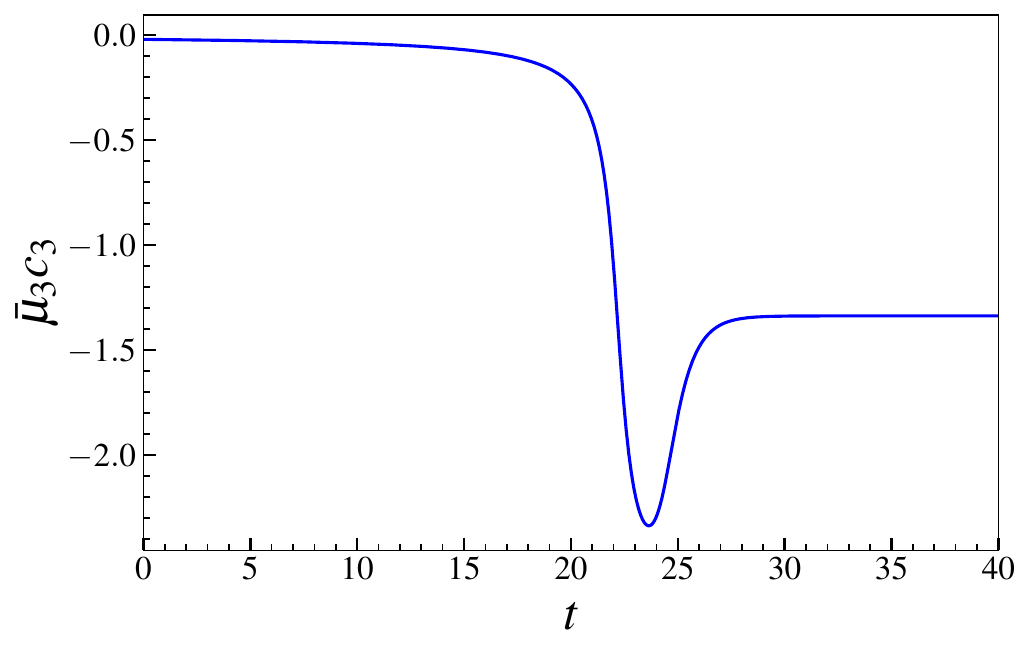}\ \ \ \ \ \ \ \ \ \ 
    \includegraphics[width=0.43\linewidth]{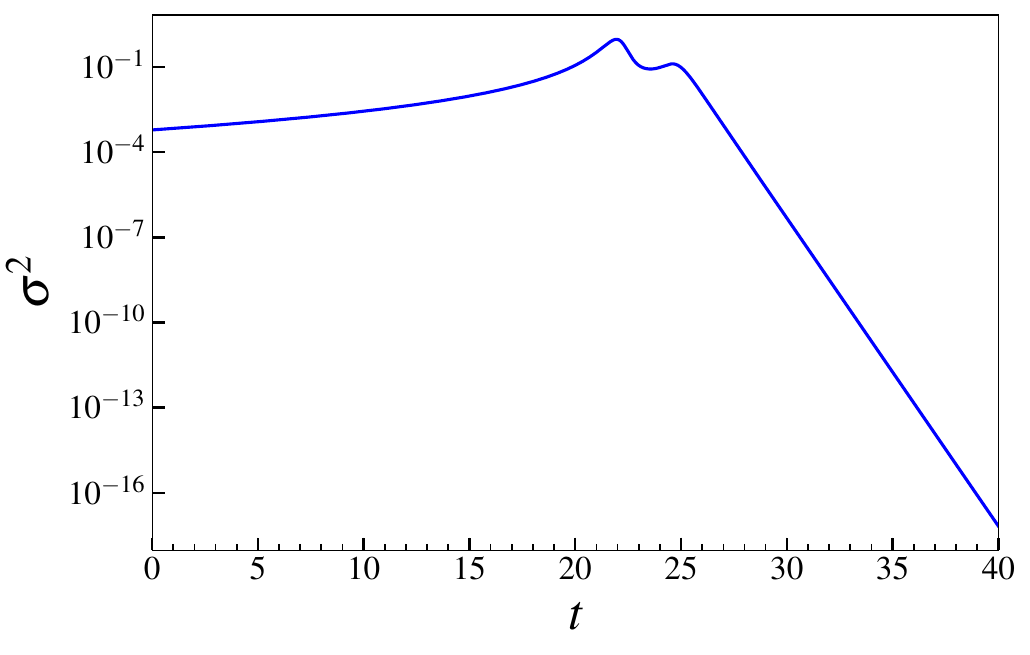}
    \includegraphics[width=0.43\linewidth]{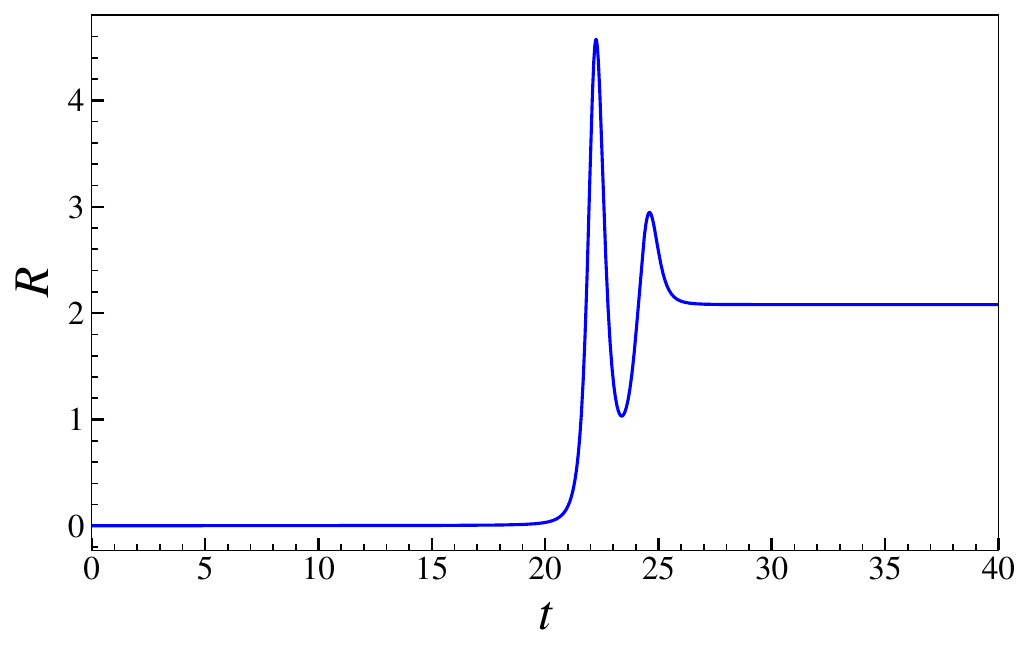}\ \ \ \ \ \ \ \ \ \ 
    \includegraphics[width=0.43\linewidth]{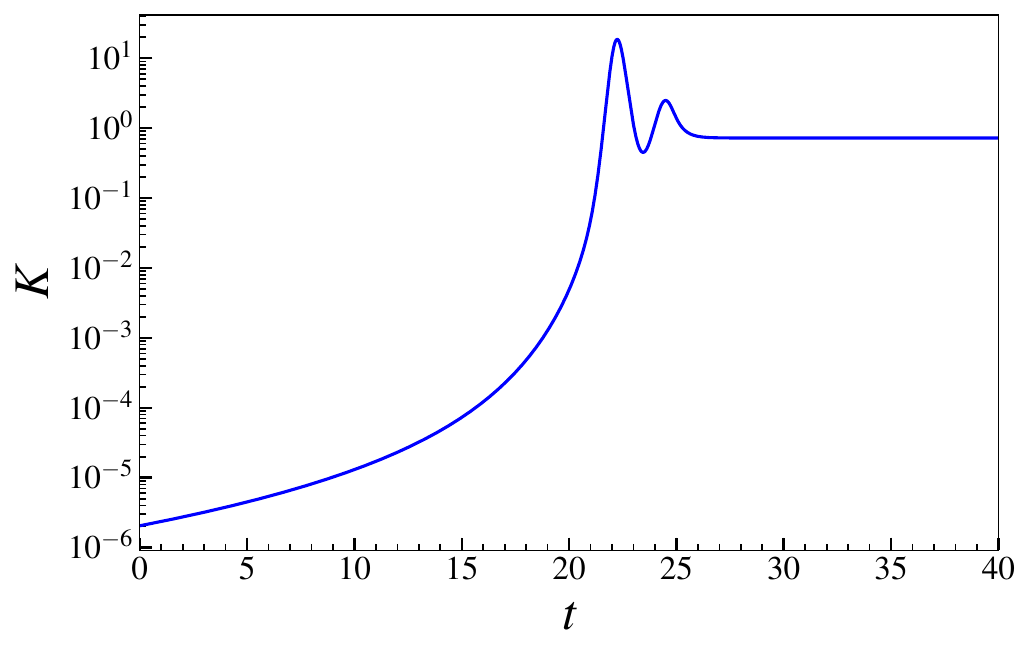}
    \caption{The time evolution of $\bar \mu_{i}c_{i}$, $\sigma^2$, Ricci scalar $R$, and Kretschmann scalar $K = R_{abcd}R^{abcd}$ is computed for initial conditions $p_{1}=1000$, $p_{2}=2000$, $p_{3}=3000$, $c_{1} = -0.13$, $c_{2}=-0.12$, and dust field energy density $\rho_{m} = 3.55 \times 10^{-5}$. There is an emergent Planckian cosmological constant in the post-bounce branch, as a result of which the universe becomes isotropic, behaving like a de Sitter universe, but remains in the quantum regime with no graceful exit to the classical universe even in the macroscopic regime.}
\label{muci-dust}
\end{figure}

To clarify the underlying physics behind the quantum gravitational isotropization mechanism identified in Bianchi-I version of mLQC-I, we solve effective Hamilton's equations with the same initial conditions chosen in Ref. \cite{Gan:2025uvt} for a dust field. In Fig. \ref{dust}, we plot the time evolution of the directional scale factors and the mean scale factor computed for the parameters $p_{1}=1000$, $p_{2}=2000$, $p_{3}=3000$, $c_{1}=-0.13$, $c_{2}=-0.12$, and dust field energy density $\rho_{m}=3.55 \times 10^{-5}$ (all values are in Planck units). As is clear from the behavior of the mean scale factor, the universe starts in the contracting branch, where all three directional scale factors contract. Then the universe bounces to the post-bounce branch, where all three directional scale factors expand almost at the same rate. This implies that the universe becomes isotropic in the post-bounce branch in the late-time regime far away from the bounce. In Ref. \cite{Gan:2025uvt}, the authors present this as one of their main results, arguing that quantum gravitational effects can isotropize the universe without invoking inflation or ekpyrotic scenario. This is a particularly interesting conclusion if the universe becomes classical when isotropized. However, as we discuss below, a more careful analysis indicates that the above assumption does not hold. 

To illuminate this, in Fig. \ref{muci-dust}, we plot the time evolution of $\bar \mu_{i}c_{i}$, $\sigma^2$, Ricci scalar $R$, and Kretschmann scalar $K = R_{abcd}R^{abcd}$ computed for the same initial conditions considered in Fig. \ref{dust}. From plots for $\bar{\mu}_{i} c_{i}$, we observe that although these variables begin with a value close to zero in the pre-bounce branch, all three go to a value almost equal to $-\pi/2$. This implies that a large classical contracting universe bounces to a large expanding universe, which remains in the quantum regime with no exit to the classical universe even in the macroscopic limit. From plots for the Ricci scalar and Kretschmann scalar, one finds that though the dynamical evolution in the pre-bounce branch corresponds to a universe filled with dust, quantum gravitational effects produce an effective Planckian cosmological constant in the post-bounce branch, which subdues the dust field. Hence, the universe behaves as a de Sitter universe in the post-bounce branch. These results reveal two important features of the quantum gravitational isotropization mechanism observed in anisotropic version of mLQC-I: first, such a mechanism arises due to the emergence of a de Sitter universe in the post-bounce branch; second, the universe remains in the quantum regime even if it has macroscopic size due to the Planckian value of the emergent cosmological constant. Recall that the emergence of the Planckian cosmological constant and the lack of exit to the classical universe are also observed in the isotropic mLQC-I, which is a limit of its anisotropic generalization. Hence, the emergence of such features in the anisotropic model is naturally expected. This implies that the quantum gravitational isotropization mechanism identified in anisotropic generalization of mLQC-I comes with a steep price: it does not allow an exit to a classical universe.

\begin{figure}
    \centering
    \includegraphics[width=0.47\linewidth]{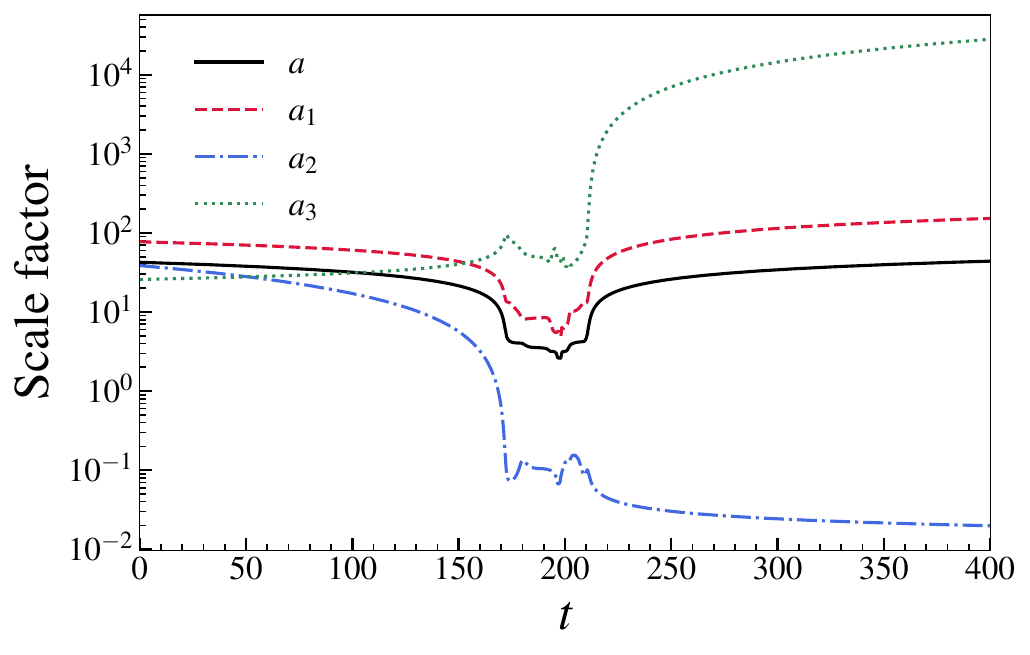}
    \caption{The time evolution of the directional scale factors and mean scale factor is computed for initial conditions  $p_{1}=1000$, $p_{2}=2000$, $p_{3}= 3000$, $c_{1} = -0.03$, and $c_{2}=-0.05$ in vacuum spacetime. The universe remains anisotropic, as there is no emergent Planckian de Sitter universe in the post-bounce branch.}
\label{vacuum}
\end{figure}

\begin{figure}
    \centering
    \includegraphics[width=0.43\linewidth]{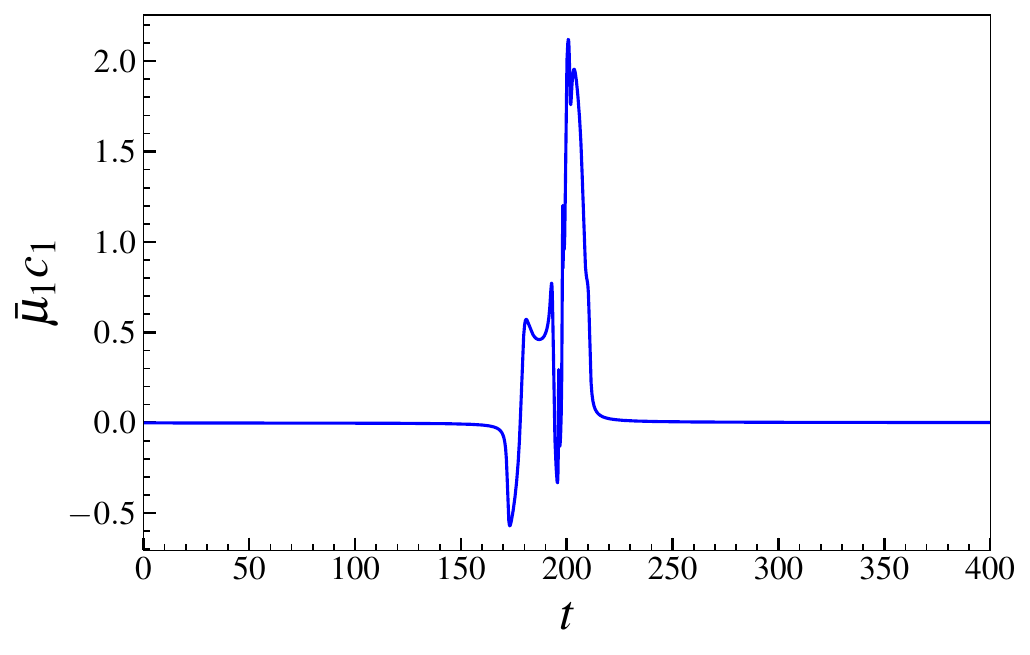} \ \ \ \ \ \ \ \ \ \ 
    \includegraphics[width=0.43\linewidth]{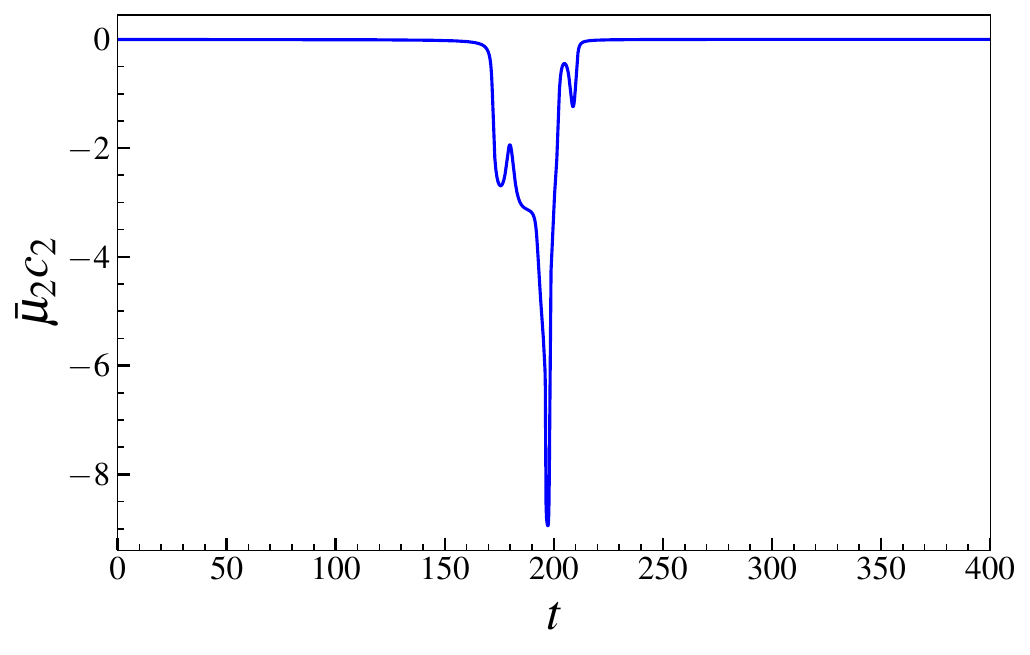}
    \includegraphics[width=0.43\linewidth]{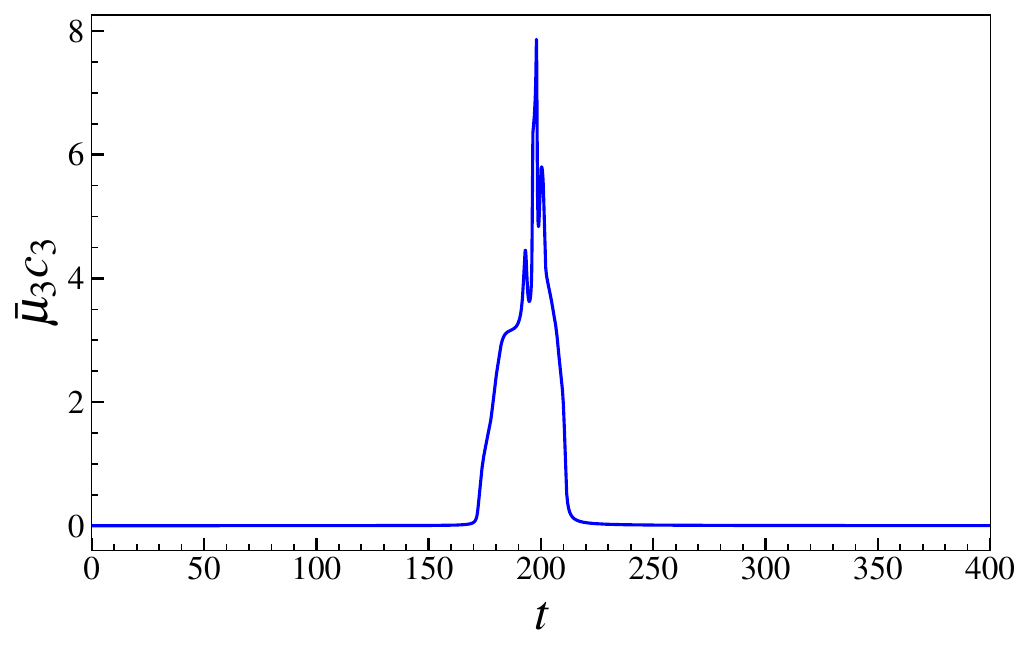}\ \ \ \ \ \ \ \ \ \ 
    \includegraphics[width=0.43\linewidth]{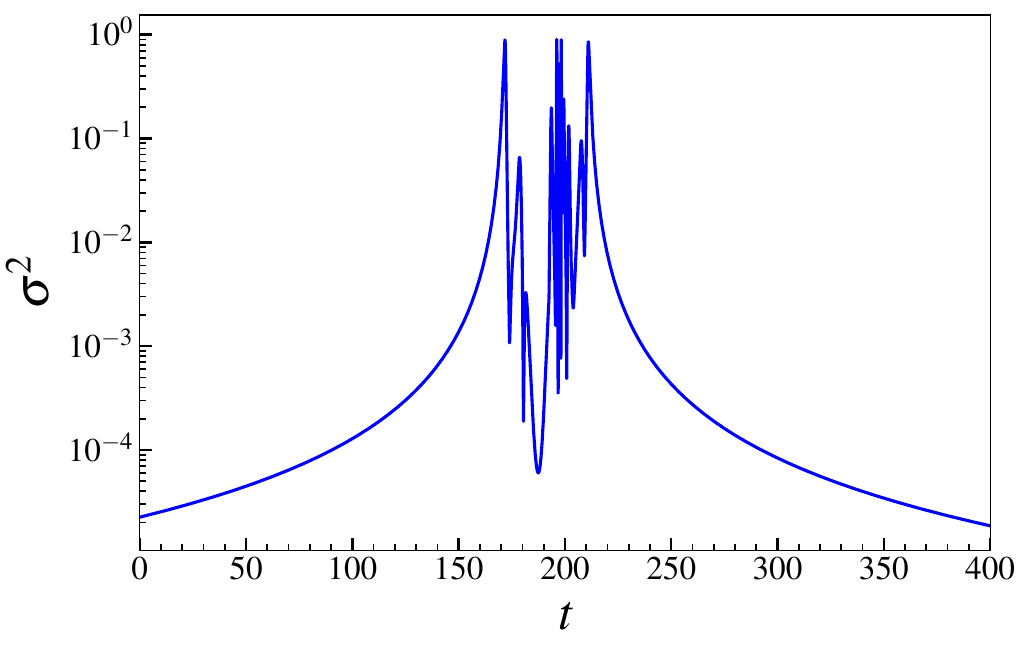}
    \includegraphics[width=0.43\linewidth]{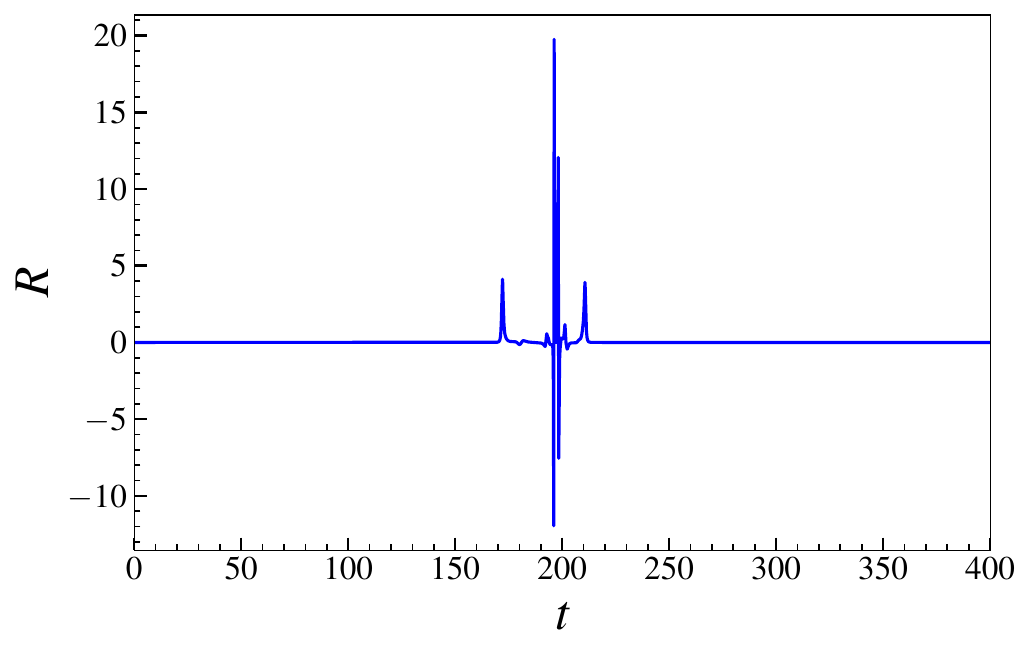}\ \ \ \ \ \ \ \ \ \ 
    \includegraphics[width=0.43\linewidth]{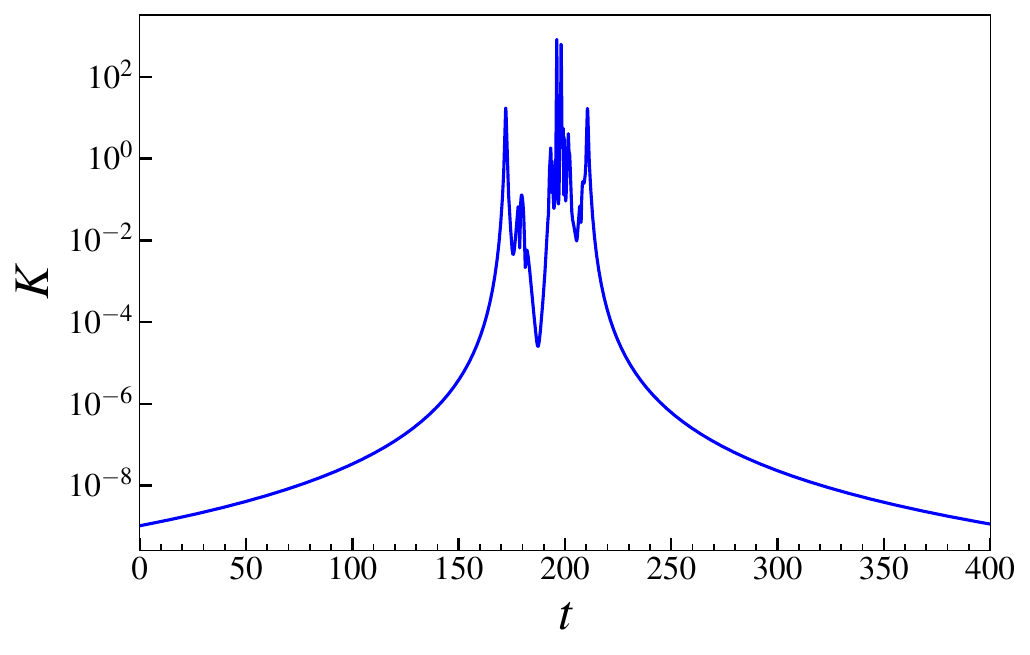}
    \caption{The time evolution of $\bar \mu_{i}c_{i}$, $\sigma^2$, Ricci scalar $R$, and Kretschmann scalar $K = R_{abcd}R^{abcd}$ is computed for initial conditions $p_{1}=1000$, $p_{2}=2000$, $p_{3}=3000$, $c_{1}=-0.03$, and $c_{2}=-0.05$ in vacuum spacetime. The universe becomes classical in the post-bounce branch, as there is no emergent Planckian de Sitter universe in the post-bounce branch.}
\label{muci-vacuum}
\end{figure}

\subsection{Non-genericness of Isotropization Mechanism}

The problem of having no graceful exit to the classical universe already shows the limitations of  the quantum gravitational isotropization mechanism identified in anisotropic version of mLQC-I. It is nevertheless insightful to understand whether such a mechanism is generic. In this pursuit, we perform several numerical simulations for a range of initial conditions for Bianchi-I vacuum spacetime and also for a Bianchi-I universe sourced individually with dust, radiation, and a massless scalar field. As an illustrative example, in Fig. \ref{vacuum}, we plot the time evolution of the directional scale factors and mean scale factor computed for the parameters $p_{1}=1000$, $p_{2}=2000$, $p_{3}=3000$, $c_{1}=-0.03$, and $c_{2}=-0.05$ in the Bianchi-I vacuum spacetime. It is obvious that the universe again begins with two scale factors contracting, while the other one expands. Then, all scale factors bounce, and two scale factors expand while the third one contracts. Therefore, from the evolution of directional scale factors, one then finds that an anisotropic universe in the pre-bounce branch remains anisotropic in the post-bounce branch with no isotropization mechanism. 

\begin{figure}[t!]
    \centering
    \includegraphics[width=0.47\linewidth]{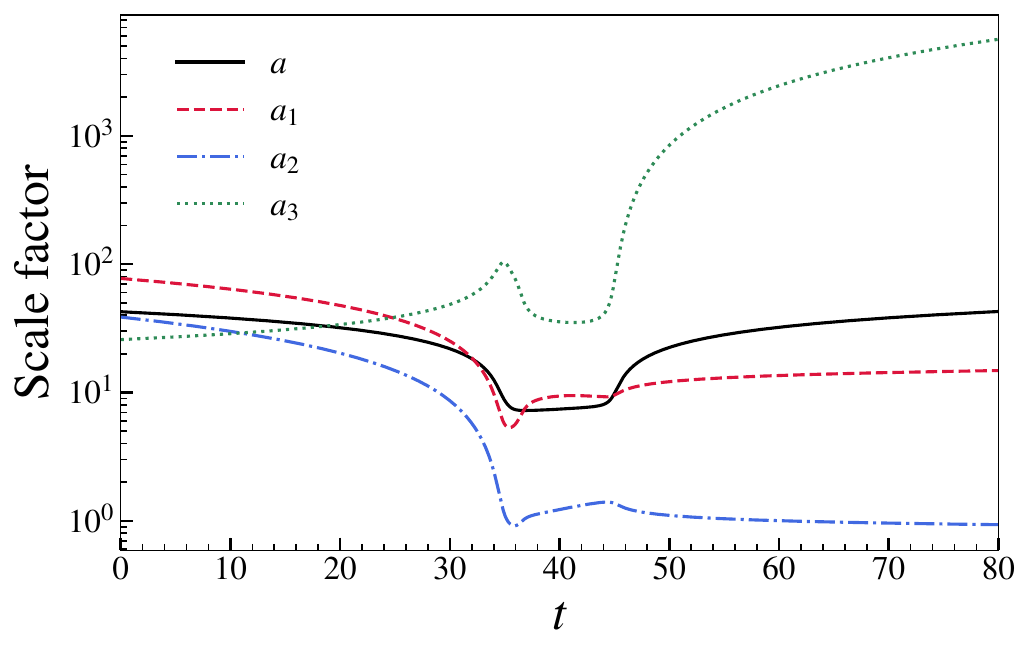}
    \caption{The time evolution of the directional scale factors and mean scale factor is computed for the initial conditions $p_{1}=1000$, $p_{2}=2000$, $p_{3}=3000$, $c_{1}=-0.3$, $c_{2}=-0.2$, and radiation field energy density $\rho_{r}=5 \times 10^{-7}$. The universe remains anisotropic, as there is no emergent Planckian de Sitter universe in the post-bounce branch for this set of initial conditions.}
\label{radiation}
\end{figure}

To understand the reason for the lack of isotropization and also the behavior of the universe in the post-bounce branch (either remaining in the quantum regime or becoming classical), in Fig. \ref{muci-vacuum}, the time evolution of $\bar \mu_{i}c_{i}$, $\sigma^2$, the Ricci scalar, and the Kretschmann scalar are computed for the same initial conditions considered in Fig. \ref{vacuum} in vacuum spacetime. As is obvious from the $\bar\mu_{i}c_{i}$ plots, all three start from a value close to zero in the pre-bounce branch and again reach a zero value in the post-bounce branch. This ensures that a large classical contracting universe bounces to a large classical expanding universe, as we expect in LQC. Consequently, one can observe that the value of anisotropic shear at the same volume remains the same in pre-bounce and post-bounce branches, recovering the conservation of anisotropic shear in the classical Bianchi-I spacetimes \cite{Motaharfar:2023hil}. Moreover, the plots for curvature invariants, namely the Ricci and Kretschmann scalars, demonstrate that the universe has low curvature values in the pre-bounce and post-bounce branches, ensuring that the quantum effects are negligible in the large volume regime. Interestingly, our results show that there is no quantum gravitational isotropization mechanism in the post-bounce branch for vacuum Bianchi-I spacetime in the anisotropic mLQC-I due to the lack of an emergent cosmological constant in the post-bounce branch. This is an interesting finding, as it points to an interplay between matter and anisotropies that produces such an effective cosmological constant and is absent in the case of a vacuum spacetime.

\begin{figure}[t!]
    \centering
    \includegraphics[width=0.43\linewidth]{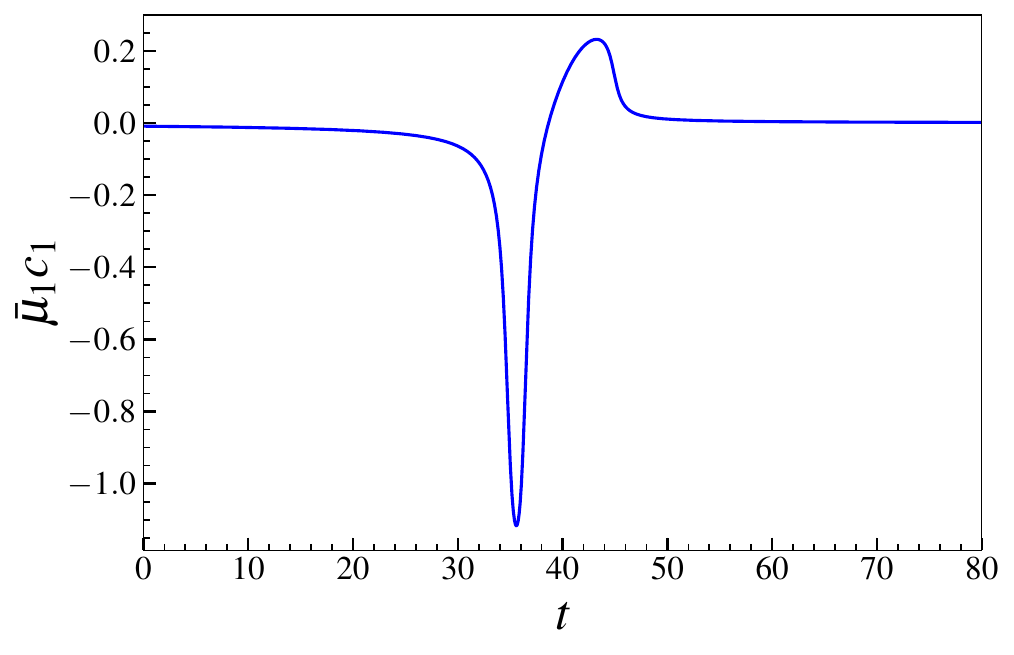} \ \ \ \ \ \ \ \ \ \ 
    \includegraphics[width=0.43\linewidth]{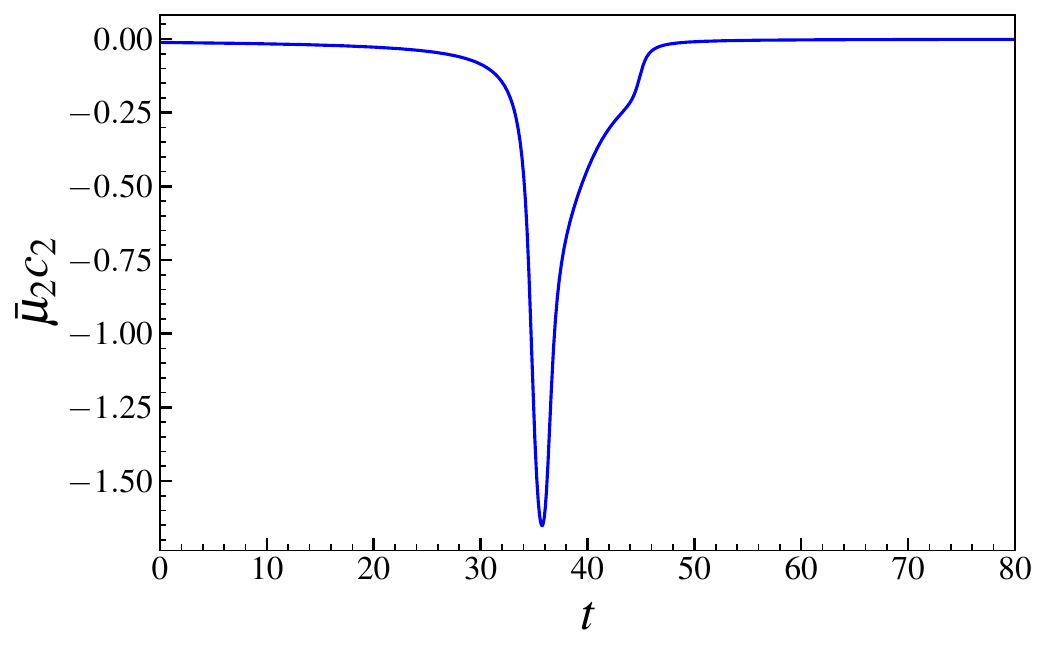}
    \includegraphics[width=0.43\linewidth]{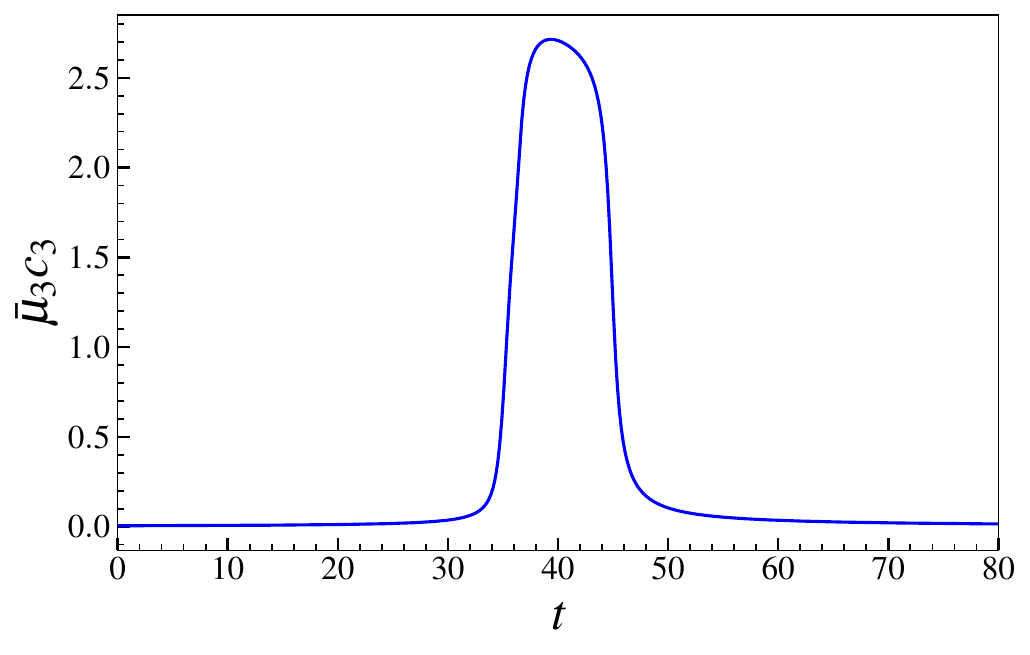}\ \ \ \ \ \ \ \ \ \ 
    \includegraphics[width=0.43\linewidth]{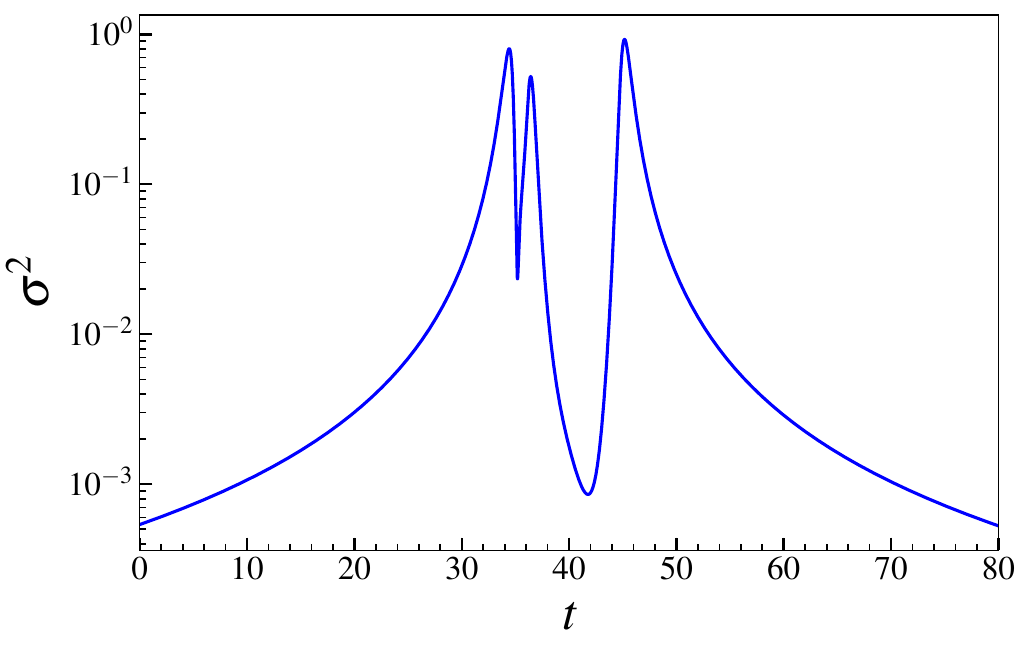}
    \includegraphics[width=0.43\linewidth]{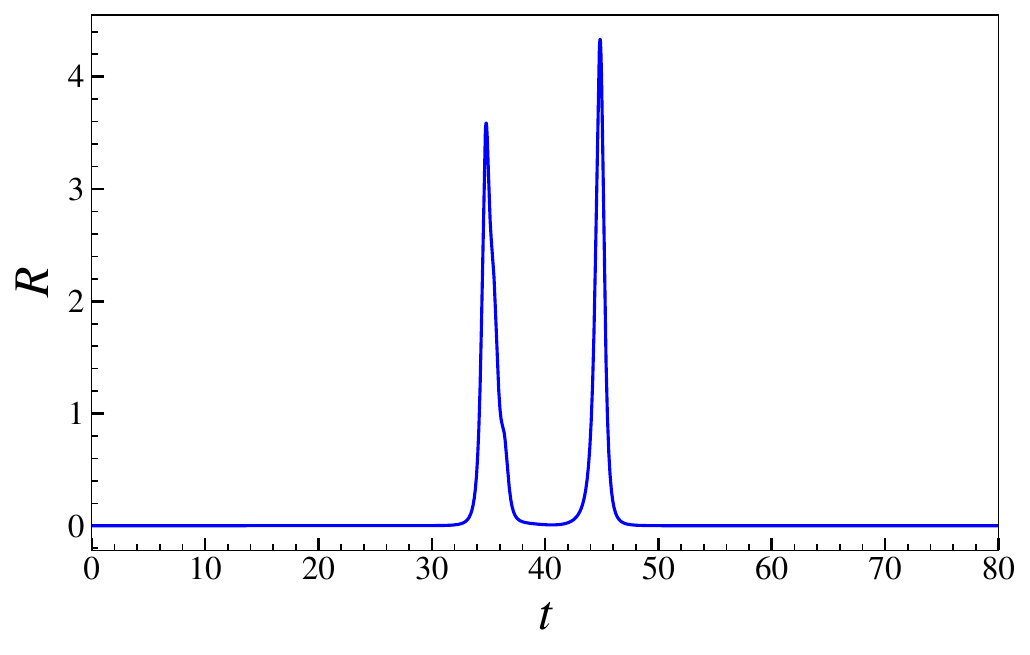}\ \ \ \ \ \ \ \ \ \ 
    \includegraphics[width=0.43\linewidth]{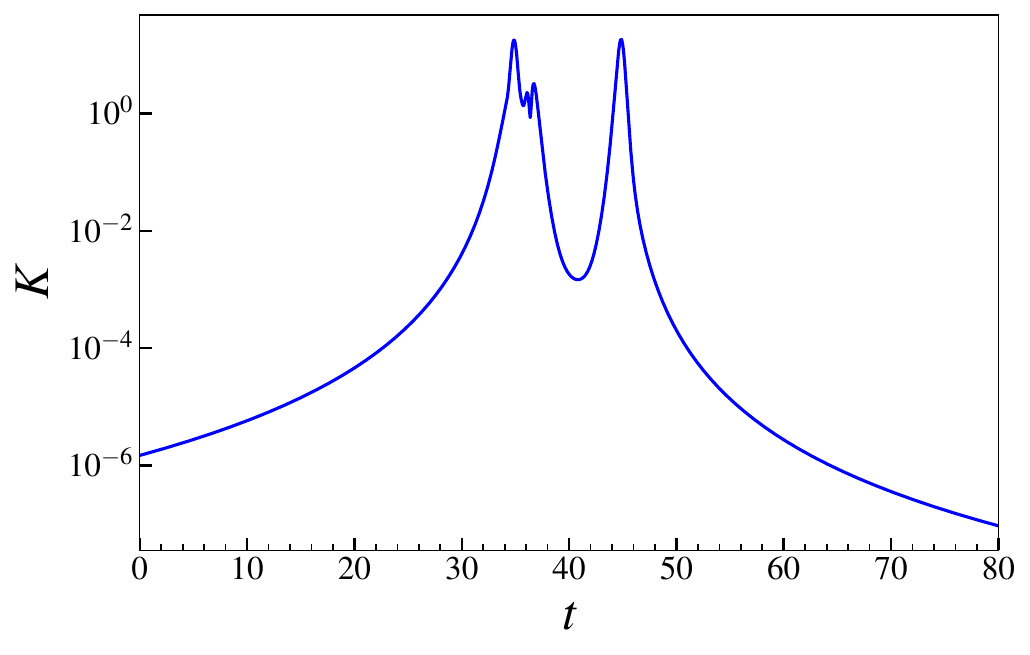}
    \caption{The time evolution of $\bar \mu_{i}c_{i}$, $\sigma^2$, Ricci scalar $R$, and Kretschmann scalar $K = R_{abcd}R^{abcd}$ is computed for the initial conditions $p_{1}=1000$, $p_{2}=2000$, $p_{3}=3000$, $c_{1} = -0.3$, $c_{2}=-0.2$, and matter density $\rho_{r} = 5 \times 10^{-7}$. The universe becomes classical in the post-bounce branch, as there is no emergent Planckian de Sitter universe in the post-bounce branch for this set of initial conditions.}
\label{muci-radiation}
\end{figure}

In order to check the robustness of our results, which demonstrate the lack of isotropization, we also perform several numerical simulations with dust, radiation, and a massless scalar field and for different ranges of initial conditions. As an illustrative example, in Fig. \ref{radiation}, we plot the time evolution of the directional scale factors and mean scale factor computed for the parameters $p_{1}=1000$, $p_{2}=2000$, $p_{3}=3000$, $c_{1} = -0.3$, $c_{2}=-0.2$, and radiation field energy density $\rho_{r} = 5 \times 10^{-7}$. From this plot one can again observe that the universe remains anisotropic in the post-bounce branch for this set of initial conditions. As a result, there is no quantum gravitational isotropization mechanism in this case similar to the vacuum spacetime. We also plot the time evolution of $\bar \mu_{i}c_{i}$, $\sigma^2$, Ricci scalar, and Kretschmann scalar for the same initial conditions used in Fig. \ref{muci-radiation}. From these plots, one can see that the anisotropic shear remains the same at the same volume in the pre-bounce and post-bounce regimes at macroscopic sizes, similar to the case of the vacuum spacetime. In addition, the curvature invariants also become negligible in the macroscopic regime. This indicates that the model recovers the classical GR in this regime. Similarly, for dust and massless scalar fields, we also find examples that the isotropization mechanism is not generic, and for a range of parameters, the universe remains anisotropic in the post-bounce branch.

\textcolor{black}{A common finding from our numerical studies is that when the initial triads and connections are held fixed and the energy density is decreased, no isotropization occurs. This might be related to the findings in Ref. \cite{Gan:2026ivx}, in which they incorrectly concluded that such an isotropization mechanism occurs only for a set of physically admissible initial conditions (point-like evolutions rather than cigar-like evolutions considered here). However, as we have already discussed in introduction, the evolution of the universe can be either point-like or cigar-like for a classical contracting universe. Depending on the interplay of matter content and anisotropic shear, either type can be more prevalent. When no matter content dominates over the anisotropic shear, the evolution is cigar-like. Therefore, the initial conditions considered here are not only physically admissible but also more prevalent than point-like initial conditions. This strengthens our conclusions that the isotropization mechanism in Bianchi-I mLQC-I is non-generic, occurring only for finely tuned point-like evolutions considered in Ref. \cite{Gan:2026ivx}. Further numerical investigation is needed to elucidate the connection between the matter-anisotropy interplay and the emergence of the cosmological constant. Taken together, our results show that the quantum gravitational isotropization mechanism observed in Bianchi-I mLQC-I does not occur for vacuum spacetime and for a range of physically admissible initial conditions in the case of a universe filled with a perfect fluid. This implies that the isotropization mechanism not only lacks a graceful exit to the classical universe but also is non-generic for all types of matter content and ranges of initial conditions. Therefore, the question of whether a robust quantum gravitational isotropization mechanism exists remains open.}

\section{Conclusions}\label{Section V}

In this work, we revisited the Bianchi-I spacetimes for a modified LQC framework motivated by Thiemann's regularization of the Hamiltonian constraint to understand the underlying physics of the quantum gravitational isotropization mechanism studied earlier \cite{Gan:2025uvt}. We performed numerical simulations for various choices of matter as well as the vacuum spacetime. For certain initial conditions and matter content, we reconfirmed that the anisotropic universe in the pre-bounce branch becomes isotropic in the post-bounce branch as all three factors exponentially expand. However, analyzing the classicality condition, i.e., $\bar \mu_{i}c_{i} \sim n\pi$, we found that although all of them are close to zero in the pre-bounce branch, none of them goes to the $n \pi$ value in the post-bounce branch. This implies that the large contracting classical universe bounces to a large expanding isotropic quantum universe with no graceful exit to a classical universe. Computing the curvature invariants, namely the Ricci scalar and the Kretschmann scalar, we found that they both approach a Planckian constant value in the post-bounce branch. This shows that the quantum gravitational isotropization mechanism occurs due to the existence of an emergent de Sitter universe in the post-bounce branch, while its Planckian value explains why the universe does not become classical even if it has a macroscopic size. Such an emergent Planckian cosmological constant was expected, as it has already been observed in isotropic mLQC-I, which is a limit of Bianchi-I version of mLQC-I when anisotropic shear scalar vanishes. In the isotropic model, the initial conditions are fixed such that the de Sitter phase appears in the pre-bounce contracting branch; as a result, the universe becomes classical in the expanding post-bounce branch. Similarly, one can enforce the emergence of the Planckian cosmological constant in the pre-bounce branch of the anisotropic model; then the universe would become classical in the post-bounce branch, while it remains anisotropic. \textcolor{black}{
Within the homogeneous effective isotropizing branch analyzed here, post-bounce classicality is not attained.
Therefore, choosing the isotropization mechanism for the universe, which amounts to clearing the anisotropic hair, has a steep price, as the universe remains in a quantum regime with no exit to the classical universe within the homogeneous effective dynamics studied here. In this regard, it is important to note that Ref. \cite{Gan:2025uvt} does not establish such an exit through backreaction in anisotropic mLQC-I. Rather, after completing its homogeneous effective analysis, it only invokes infrared backreaction of super-Hubble modes, by analogy with the broader de Sitter literature, as a possible mechanism for relaxing the post-bounce accelerated phase. Such a mechanism has not been demonstrated for the anisotropic mLQC-I solutions considered here.}

\textcolor{black}{In this regard, it is important to distinguish the quantum-gravitational isotropization mechanism analyzed in the present work from the standard asymptotic behavior of classical expanding Bianchi-I solutions. In the isotropizing solutions studied here, the suppression of anisotropy is tied to the emergence of a post-bounce Planckian de Sitter phase. For this reason, the resulting macroscopic universe does not become classical in the post-bounce branch. By contrast, if the emergent Planckian cosmological constant is realized in the pre-bounce branch, then the pre-bounce branch is quantum while the post-bounce branch may become classical. Since in classical Bianchi-I dynamics the shear scalar decays as $\sigma^2 = \Sigma^2/a^6$, whereas the corresponding matter energy density decays more slowly with expansion, anisotropies can become subdominant for matter such as dust or radiation in the expanding branch, but only at very late times. This is simply the standard asymptotic behavior of classical GR and is therefore conceptually distinct from the de Sitter-driven isotropization discussed in this paper, which occurs in the deep quantum regime before classicality is restored.}

An interesting question is to explore the genericness of  isotropization mechanism, in particular for a universe filled with a perfect fluid. Contrary to earlier expectations \cite{Gan:2025uvt}, we find that for some range of initial conditions, the universe filled with a perfect fluid does not isotropize in the post-branch universe. Interestingly, it becomes classical in these cases due to the absence of the Planckian de Sitter universe in the post-bounce branch. These results show that the isotropization mechanism is not generic. Furthermore, we also found that in the case of the vacuum spacetime, the universe is anisotropic both in the pre-bounce and post-bounce branches while it becomes classical in the large volume regime. In fact, this implies that the quantum gravitational isotropization mechanism can occur only in case the universe is filled with matter content. This result demonstrates further limitations of the genericness of such an isotropization mechanism.

In summary, our analysis shows that the quantum gravitational isotropization mechanism identified in Bianchi-I generalization of mLQC-I faces two key limitations: first, in the post-bounce branch it does not provide a graceful exit to a classical universe; second, it is non-generic, failing for a range of admissible initial conditions. Our analysis also demonstrates that the macroscopic loop quantum universe does not by itself guarantee classicality, in agreement with the findings in Ref. \cite{Motaharfar:2023hil}. The results highlight the need for a careful examination of the behavior of connection components and curvature invariants in both the pre-bounce and post-bounce regimes in order to characterize the classical properties of spacetime. This exercise clarifies that the anisotropic shear damping mechanism in the anisotropic generalization of mLQC-I  is neither mysterious nor generic and arises quite naturally from the Planckian de Sitter character of the post-bounce universe. Moreover, our results indicate that, unlike in the isotropic model, there exists a range of initial conditions for which mLQC-I with anisotropies yields a classical universe on both sides of the bounce, a feature that merits further detailed investigation. Finally, it is important to note that our conclusions apply only to Bianchi-I version of mLQC-I and not to all Thiemann regularized loop quantum cosmologies. In fact there exist versions of Thiemann regularized LQC, such as mLQC-II where the bounce is symmetric in the isotropic setting as in standard LQC. In this setting there is no emergence of a Planckian cosmological constant and the classical limit is obtained in the macroscopic regime \cite{Li:2021mop}. One may therefore expect that an anisotropic generalization of mLQC-II will be free from the limitations of Bianchi-I mLQC-I discussed in this work, and it will be worthwhile to investigate this in future studies. 

\textcolor{black}{
Ref. \cite{Gan:2026ivx} recently questioned the physical admissibility of the initial conditions used in our simulations. However, classical Kasner behavior already shows that, as a contracting Bianchi-I universe approaches the singularity, the evolution can generically be either point-like, with all three scale factors contracting, or cigar-like, with two scale factors contracting and one expanding. Which of these behaviors is realized depends on the relative strength of matter and anisotropic shear. In particular, when matter does not dominate over the shear, the approach to the singularity is generically cigar-like. The initial conditions used in this work are therefore not only physically admissible, but in this regime they are in fact the more natural and prevalent class. Consequently, the claim of generic isotropization cannot be maintained by restricting attention a posteriori to a finely tuned subclass of point-like initial data. Our analysis instead shows that isotropization in mLQC-I is non-generic and arises only for a special class of point-like evolutions considered in Refs. \cite{Gan:2025uvt,Gan:2026ivx}.}

\section*{Acknowledgments}
 
P.S. thanks Jorge Pullin and Anzhong Wang for discussions. This work is supported by NSF grants PHY-2206557 and PHY-2409543.

\end{document}